\documentclass[aps,prl,showpacs,groupedaddress]{revtex4}  % for review and submission
\usepackage{graphicx}  % needed for figures
\usepackage{dcolumn}   % needed for some tables
\usepackage{bm}        % for math
\usepackage{amssymb}   % for math

\begin{document}

% The following information is for internal review, please remove them for submission
%\widetext
%\leftline{Version xx as of \today}
%\leftline{Primary authors: Joe E. Physics}
%\leftline{To be submitted to (PRL, PRD-RC, PRD, PLB; choose one.)}
%\leftline{Comment to {\tt d0-run2eb-nnn@fnal.gov} by xxx, yyy}

% the following line is for submission, including submission to the arXiv!!
%\hspace{5.2in} \mbox{Fermilab-Pub-04/xxx-E}

\title{Modeling high-energy cosmic ray induced terrestrial muon flux: A lookup table}
                             % of this file prior to submission, they
                             % contain a time stamp for the authorlist)
                             % (includes institutions and visitors)
\author{Dimitra Atri*}
\author{Adrian L. Melott}
\affiliation{Department of Physics and Astronomy, University of Kansas, 1251 Wescoe Dr. \# 1082, Lawrence, KS 66045, United States of America, *Email: dimitra@ku.edu}

\begin{abstract}
On geological timescales, the Earth is likely to be exposed to an increased flux of high energy cosmic rays (HECRs) from astrophysical sources such as nearby supernovae, gamma ray bursts or by galactic shocks. Typical cosmic ray energies may be much higher than the $\leq$ 1 $GeV$ flux which normally dominates. These high-energy particles strike the Earth's atmosphere initiating an extensive air shower. As the air shower propagates deeper, it ionizes the atmosphere by producing charged secondary particles. Secondary particles such as muons and thermal neutrons produced as a result of nuclear interactions are able to reach the ground, enhancing the radiation dose. Muons contribute 85\% to the radiation dose from cosmic rays. This enhanced dose could be potentially harmful to the biosphere. This mechanism has been discussed extensively in literature but has never been quantified. Here, we have developed a lookup table that can be used to quantify this effect by modeling terrestrial muon flux from any arbitrary cosmic ray spectra with 10 $GeV$ - 1 $PeV$ primaries. This will enable us to compute the radiation dose on terrestrial planetary surfaces from a number of astrophysical sources.
\end{abstract}

\maketitle

%\section{\label{sec:level1}First-level heading}

% sections are not used for PRL papers
\section{1. Introduction}
There is a non-trivial probability of high-energy astrophysical events such as nearby supernovae (Fields et al., 2008) and gamma ray bursts (Dermer and Holmes, 2005; Dermer, 2010; Kusenko, 2010) (GRBs) to expose the Earth to an enhanced flux of radiation over Gyr timescales. Along with higher flux, the peak of energy spectrum of primaries is also shifted toward higher energies. This radiation consists of both photons and charged nuclei, also known as cosmic rays. Periodic motion of the Sun perpendicular to the galactic plane can also expose us to an enhanced level of HECRs generated by the galactic shock (Medvedev and Melott, 2007). Detailed modeling of the effects of photons hitting the EarthÕs atmosphere exists (Thomas et al., 2005) and will not be discussed further. We focus on the effects of the cosmic ray component of this radiation in this paper, which are of primary importance to the biosphere. Cosmic ray primaries interact with the nuclei of the atmosphere, initiating a series of reactions propagating deeper into the atmosphere, known as extensive air showers. The shower propagates down towards the ground with the shower-front shaped like a pancake. Most of the energy of the primary particle is used up in ionizing the atmosphere. This component ionizing the atmosphere consists of charged particles and photons, also referred to as the electromagnetic component of the shower. This electromagnetic component is mostly absorbed within first 1000 $g$ $cm^{-2}$ of atmosphere and mostly the hard component (muons) of the shower is able to reach the ground. Other particles and photons reaching the ground are not biologically effective.  This atmospheric ionization induced by HECR primaries has already been modeled for energies up to 1 PeV (Atri et al., 2010). Along with ionizing the atmosphere, secondary particles reach ground level. Depending on their energy, they have the ability to penetrate deep underground and into water. Enhanced levels of secondary particle flux could be harmful to living organisms (Dar et al., 1998) by increased mutation rates, cancer and birth defects (Juckett, 2009).

Muons are generated by the decay of charged pions and have a very small interaction cross section. Also, their cross section is a very weak function of energy and the ionization energy loss is a very small fraction of the primary energy, even at very high energies.  As a result, high energy muons hit the ground practically unhindered, generated by high energy primaries. But since muons are charged particles, they are easy to detect and give a dominant signal deep in the atmosphere and underground. They are usually detected with detectors buried under a meter or more underground in order to absorb the electromagnetic component.

The muon component dominates the flux of particles on the ground at energies above 100 $MeV$. Its contribution is about 85\% of the total equivalent biological dose by cosmic rays (Alpen, 1998). At lower energies, the muon flux decreases because of their short lifetime of 2 $\mu$s. Muons with higher energy travel with higher velocities and hence travel longer distances due to the time dilation effects. They lose energy due to ionization at a rate of about 2 $MeV$ per $g$ $cm^{-2}$. Since the column density of our atmosphere is about 1000 $g$ $cm^{-2}$, a muon loses around 2 $GeV$ of energy, on an average, upon reaching the ground (falling vertically). This radiation dose from muons at the ground level will be in addition to the dose from other components discussed elsewhere (Karam, 1999; Karam, 2002a,b).

Analytical models (Gaisser, 1990) can approximate muon flux resulting from vertical GCR primaries and vast amount of data (Hebbeker, 2002) is available from detectors around the world. But no model is available to compute muon flux when the CR primary flux is different from the normal GCR flux. In particular at higher energies, the hadronic interaction cross sections are only estimates, and reasonable data can be obtained only by computational methods.

We have developed a lookup table to compute terrestrial flux of muons resulting from high energy cosmic rays interaction with the atmosphere that can be used by convolving with the spectrum from any astrophysical source. This work is complementary to the already existing methods (Gaisser, 2002; Agrawal et al., 1996; Hebbeker et al., 2002) for evaluating the muon flux from primary cosmic ray spectrum.
\section{2. Computational Modeling}

We use CORSIKA (COsmic Ray SImulations for KAscade) (Pierog et al.) which is a Monte Carlo code used extensively to study air showers generated by primaries up to 100 $EeV$. It is primarily calibrated by KASCADE data, which is a detector used to study hadronic interactions in the $10^{16}$ to $10^{18}$ $eV$ energy range. We use the code to simulate air showers for a flat spectrum of high-energy primaries (protons) so that any arbitrary CR spectra can be applied later on, to calculate the muon flux. It should be noted that several details of shower development, especially at higher energies are subjected to uncertainties and approximations and are unavoidable in this approach. 

In order to obtain data that is independent of the primary GCR spectrum and geographic location on the Earth, we simulated showers at fixed primary energies. The flux of muons at the ground level strongly depends on the zenith angle of the primary; therefore we simulated showers for a range of zenith angles. We compiled data for primaries from 10 $GeV$ up to 1 $PeV$ in 0.1 log$_{10}$ intervals. At each energy, simulations are performed at zenith angles starting from 5, 15...up to 85 degrees. We averaged 90,000 showers at each energy for 10 $GeV$ - 1 $TeV$ range, 9,000 for 1 $TeV$ - 10 $TeV$, 900 for 10 $TeV$ - 100 $TeV$ and 180 showers for 100 $TeV$ - 1 $PeV$. Overall, the data is compiled from a large ensemble of $1.9 \times 10^6$ showers.

CORSIKA 6.960 was used for all the simulations. The code was set up with EPOS as the high-energy hadronic interaction model due to its compatibility with KASCADE data.  For low energies, the FLUKA hadronic interaction model was chosen since it is the fastest and best at tracking muons in our energy range (Pierog et al.; Pierog et al., 2007) The code was installed with the SLANT option to study the longitudinal shower development. It will be used for other purposes, not discussed here. CURVED option was chosen for primaries incident at large zenith angles and UPWARD option for albedo particles. The energy cut was set at 300 $MeV$ since it is adequate to get all the muons and hadrons produced by photon interactions while saving a significant amount of computing time. Sea level was set as the observation level.

CORSIKA outputs binary files, which record data on a number of particle species hitting the ground level, which were converted to ASCII format for analysis. Each file contains momentum data of individual particles at the sea level from a single air shower. Muon momentum is extracted from each file generated by a primary at a fixed energy and fixed zenith angle. The momentum data is then compiled for all showers at a given energy and angle, and is binned in logarithmic energy bins of 0.1 GeV in log$_{10}$ intervals. It is then normalized by the total number of showers, giving the muon flux from a single primary. As a result we get the terrestrial muon spectrum of a particular primary energy at a given zenith angle. Data obtained from different angles at a given primary energy is then averaged by sin theta weight resulting in an isotropic muon spectrum at a particular primary energy. This is the data recorded in the data table. This process is repeated for all 51 primary energies from 10 $GeV$ - 1 $PeV$.

\section{3. Results}

Muons have a lifetime of 2.2 $\mu$s. As the zenith angle of the primary particle increases, so does the distance to travel and therefore a lot of muons decay before reaching the ground. 
\begin{figure}[htp]
\centering%
\includegraphics[scale=.6]{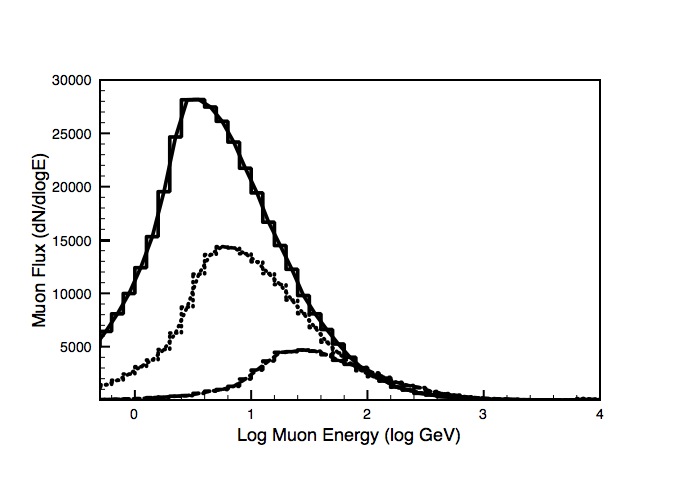}
\caption{Differential muon flux for 10 TeV primaries at zenith angles $5^{o}$ (solid), $45^{o}$ (dotted), and $85^{o}$(dashed). The flux goes down with increasing zenith angles.}
\label{figure 1}
\end{figure}

Therefore, as expected we get lesser muon flux at higher zenith angles (Figures 1 and 2). Low energy primaries are not efficient in producing muons of enough energy so that they can reach the ground. A large number of primaries at lower energies produce fewer muons at the ground level. The number of secondary particles produced in the shower increases with the primary energy and so does the number of muons hitting the ground (Figure 3). The contribution from larger zenith angle component is very small for low energy primaries, and goes up with energy. This is evident in Figure 3 where the flux from 100 GeV primary falls very steeply compared to primaries at higher energies. 

We have also compared the lookup table data with Hebbeker and Timmermans polynomial fit to data collected from a number of experiments (Hebbeker and Timmermans, 2002). We find a maximum 15\% deviation from the polynomial fit for low energy muons, 1-5\% difference in the 10 $GeV$ - 3 $TeV$ range and 8\% deviation at higher energies. Overall it is a very good fit considering the experiment to experiment variation in data compiled in the reference (Hebbeker and Timmermans, 2002). 

\begin{figure}[htp]
\centering%
\includegraphics[scale=.6]{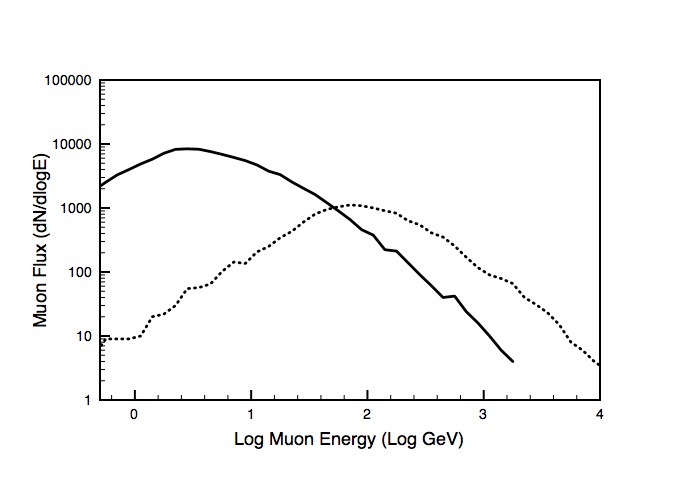}
\caption{Differential muon flux for 1 PeV primaries at zenith angles $5^{o}$ (solid) and $85^{o}$(dotted).}
\label{figure 2}
\end{figure}

 \begin{figure}[htp]
\centering%
\includegraphics[scale=.6]{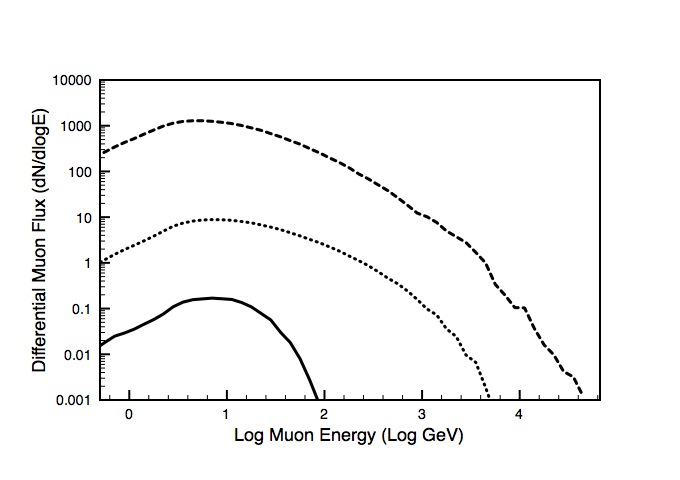}
\caption{The muon flux averaged over the hemisphere from 100 GeV (solid), 10 TeV (dotted) and 1 PeV (dashed) primaries. Flux for higher energy primaries does not fall sharply compared to the lower energy primaries since the contribution from higher zenith angles increases with increasing primary energy.}
\label{figure 3}
\end{figure}

\begin{figure}[htp]
\centering%
\includegraphics[scale=.6]{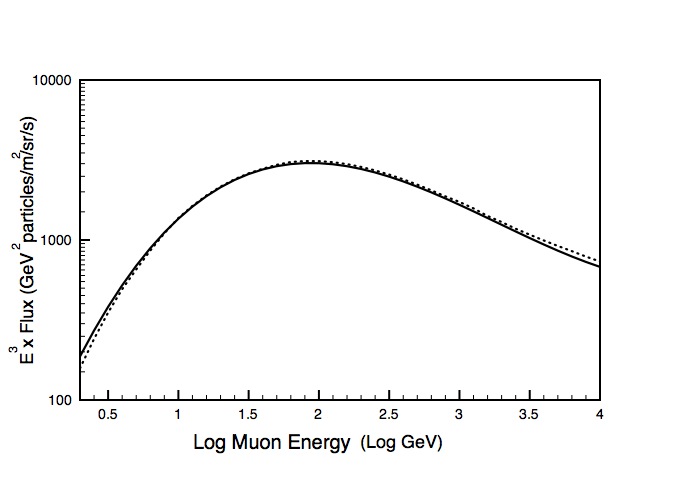}
\caption{Muon flux computed from the lookup table (dots) compared with Hebbeker and Timmermans (2002) polynomial fit (solid).}
\label{figure 4}
\end{figure}

\section{4. Using the lookup table}
Our primary energies are centered at 10 $GeV$, $10^{1.1}$ $GeV$ and so on with bin size of 0.1 in log$_{10}$ intervals. The number of particles in a given energy bin (dN/dE) can be calculated using the differential spectrum from any astrophysical source. This number is then multiplied by the data given in the table to get the corresponding muon spectrum for each primary energy. The total muon spectrum can be obtained by summing over a given energy range. Since the primaries higher than 17 $GeV$ are unaffected by the geomagnetic field, their flux is independent of the geographic location. Large amount of literature exists for muon flux for lower energies, both from experiments (Hebbeker et al., 2002) or using approximate analytical methods (Gaisser, 1990).
\section{5. Discussion}
No simple method exists to calculate the terrestrial muon flux induced by arbitrary cosmic ray spectra. We have developed a lookup table that can be used for primaries up to 1 $PeV$. The radiation dose corresponding to the increased muon flux can have significant effects on the biosphere and can be quantified using this data. Data is made freely available at: http://kusmos.phsx.ku.edu/$\sim$melott/Astrobiology.htm.
\section{Acknowledgements}
We thank Tanguy Pierog and Dieter Heck for their advice in using CORSIKA. The extensive computer simulations were supported by the NSF via TeraGrid allocations at the National Center for Supercomputing Applications, Urbana, Illinois, TG-PHY090098, TG-PHY090067T and TG-PHY090108.

\section{References}
%\begin{thebibliography}{19}
\noindent
Vivek Agrawal, Gaisser, T.K.,  Lipari, P. and Stanev, T., 1996, Atmospheric neutrino flux above 1 GeV. Phys. Rev. D 53, 1314.
\newline
Edward L. Alpen, 1998. Radiation Biophysics. Academic Press, San Diego.
\newline
Dimitra Atri, Melott, A.L. and Thomas, B.C., 2010. Lookup tables to compute high energy cosmic ray induced atmospheric ionization and changes in atmospheric chemistry. J. Cosmol. Astropart. P. 5, 008.
\newline
Arnon Dar, Laor, A. and Shaviv, N.J., 1998. Life Extinctions by Cosmic Ray Jets. Phys. Rev. Lett. 80, 26.
\newline
Charles D. Dermer and Holmes, J.M., 2005. Cosmic Rays from Gamma-Ray Bursts in the Galaxy. Astrophys. J. 628, L21.
\newline
Charles D. Dermer, 2010. First Light on GRBs with Fermi. arXiv:1008.0854.
\newline
Brian D. Fields, Athanassiadou, T. and Johnson, S.R., 2008. Supernova Collisions with the Heliosphere. Astrophys. J. 678, 549.
\newline
Thomas K. Gaisser, 1990. Cosmic rays and particle physics. Cambridge University Press, Cambridge.
\newline
Thomas K. Gaisser, 2002. Semi-analytic approximations for production of atmospheric muons and neutrinos. Astropart. Phys. 16, 285-294
\newline
Thomas Hebbeker and Timmermans, C., 2002. A compilation of high energy atmospheric muon data at sea level. Astropart. Phys. 18, 107
\newline
David A. Juckett, 2009. A 17-year oscillation in cancer mortality birth cohorts on three continents - synchrony to cosmic ray modulations one generation earlier. Int. J. Biometerol. 53, 6.
\newline
P. Andrew Karam, 1999. Calculations of Background Beta-Gamma Radiation Dose Through Geologic Time. Health Phys. 77, 6.
\newline
P. Andrew Karam, 2002a. Gamma and Neutrino Radiation Dose From Gamma Ray Bursts and Nearby Supernovae. Health Phys. 82, 4.
\newline
P. Andrew Karam, 2002b. Terrestrial radiation exposure from supernova-produced radioactivities. Radiat. Phys. Chem. 64, 2.
\newline
Alexander Kusenko, 2010. Past Galactic GRBs, and the origin and composition of ultrahigh-energy cosmic rays. arXiv:1007.096.
\newline
Mikhail V. Medvedev and Melott, A.L., 2007. Do Extragalactic Cosmic Rays Induce Cycles in Fossil Diversity? Astrophys. J. 664, 879.
\newline
Tanguy Pierog, Heck, D. and Knapp, J., CORSIKA (COsmic Ray SImulations for KAscade) an air shower simulation program homepage: \url{http://www-ik.fzk.de/corsika/}.
\newline
Tanguy Pierog, Engel, R., Heck, D., Ostapchenko, S. and  Werner, K., 2007. Prepared for 30th International Cosmic Ray Conference. Latest Results from the Air Shower Simulation Programs CORSIKA and CONEX. (ICRC 2007), Merida, Yucatan, Mexico, arXiv:0802.1262v1.
\newline
Brian C. Thomas, Melott, A.L., Jackman, C.H., Laird, C.M., Medvedev, M.V., Stolarski, R.S., Gehrels, N., Cannizzo, J.K., Hogan, D.P. and Ejzak, L.M., 2005. Gamma-Ray Bursts and the Earth: Exploration of Atmospheric, Biological, Climatic, and Biogeochemical Effects. Astrophys. J. 634, 509.

%\end{thebibliography}

\end{document}